\documentclass[journal=nalefd,manuscript=letter,layout=twocolumn]{achemso}
\usepackage{chemformula} 
\usepackage[T1]{fontenc} 
\usepackage{graphicx}
\usepackage{dcolumn}
\usepackage{bm}
\usepackage[utf8]{inputenc}
\usepackage[T1]{fontenc}
\usepackage{mathptmx}
\usepackage{etoolbox}
\usepackage{braket}
\usepackage{upgreek}
\usepackage{multirow}
\usepackage{multicol}

\makeatother

\title{Clock Transitions Generated by Defects in Silica Glass}

\author{Brendan C.~Sheehan}
\email{bcsheehan@umass.edu}
\affiliation{Department of Physics and Astronomy, Amherst College, Amherst, MA 01002, USA}
\alsoaffiliation{Department of Physics, University of Massachusetts Amherst, Amherst, MA 01003, USA}%
\author{Guanchu Chen}
\affiliation{Department of Physics and Astronomy, Amherst College, Amherst, MA 01002, USA}
\alsoaffiliation{Department of Physics, University of Massachusetts Amherst, Amherst, MA 01003, USA}%
\author{Jonathan R.~Friedman}
\affiliation{Department of Physics and Astronomy, Amherst College, Amherst, MA 01002, USA}
\alsoaffiliation{Department of Physics, University of Massachusetts Amherst, Amherst, MA 01003, USA}

\date{\today}

\begin{document}

\begin{abstract}
Clock transitions (CTs) in spin systems, which occur at avoided level crossings, enhance quantum coherence lifetimes $T_2$ because the transition becomes immune to the decohering effects of magnetic field fluctuations to first order. We present the first electron-spin resonance (ESR) characterization of CTs in certain defect-rich silica glasses, noting coherence times up to 16~$\upmu$s at the CTs. We find CT behavior at zero magnetic field in borosilicate and aluminosilicate glasses, but not in a variety of silica glasses lacking boron or aluminum. Annealing reduces or eliminates the zero-field signal. Since boron and aluminum have the same valence and are acceptors when substituted for silicon, we suggest the observed CT behavior could be generated by a spin-1 boron-vacancy center within the borosilicate glass, and similarly, an aluminum-vacancy center in the aluminosilicate glass.
\end{abstract}

\maketitle

\section{Introduction}

Quantum information science relies on qubits -- two-level systems with energy levels that can be superposed for an amount of time useful for quantum operations to be implemented. In recent years an explosion of interest in quantum information science has yielded proof-of-concept experiments and more, indicating an attractive future for quantum technologies. Commonly studied qubit technologies include superconducting circuits based on Josephson junctions~\cite{kjaergaardSuperconductingQubitsCurrent2020}, trapped ion qubits~\cite{bruzewiczTrappedionQuantumComputing2019}, defect-in-material spin qubits~\cite{mortonEmbracingQuantumLimit2011}, nuclear spin qubits~\cite{heinrichQuantumcoherentNanoscience2021}, molecular nanomagnet (MNM) systems~\cite{coronadoMolecularMagnetismChemical2019,chiesaMolecularNanomagnetsViable2024}, and hybrid architectures containing more than one type of qubit~\cite{rablHybridQuantumProcessors2006,mortonHybridSolidStateQubits2011,clerkHybridQuantumSystems2020,gimenoEnhancedMolecularSpinPhoton2020}. Because many of these systems can be manipulated and read out using spin-resonance-type techniques, electronic spin states represent viable qubit platforms for quantum information processing using electron-spin resonance (ESR).~\cite{luisObservationQuantumCoherence2000,leuenbergerQuantumComputingMolecular2001,tejadaMagneticQubitsHardware2001,friedmanSingleMoleculeNanomagnets2010} Some MNMs have have been shown to have phase memory times long enough to be useful for quantum gate operations and for storage of quantum information. 

\begin{table*}
\begin{tabular}{c||c|c|c|c|c}
 ~ &  Duran & Aluminosilicate 0812 & Vycor 7913 & AGC40 \\
 \hline
 SiO$_2$ (\%)  & 81 & 60 & 96 & 95 \\
 B$_2$O$_3$ (\%)  & 13 & 4.5 & 3 & 5 \\
 Al$_2$O$_3$ (\%)  & 2 & 14 & 0 & Trace \\
 Na$_2$O (\%)  & 4* & <0.02 & Trace & Trace \\
 K$_2$O (\%)  & 4* & 0 & 0 & 0 \\
 CaO (\%)  & 0 & 10 & 0 & 0 \\
 BaO (\%)  & 0 & 9 & 0 & 0 \\
 \hline
\end{tabular}
\caption{\textbf{Composition of each glass sample studied, as percentage by mass.} All data was sourced from commercially available descriptions of the material from the supplier. The asterisk (*) for Duran indicates that Schott provides the measure of 4\% by mass for the combination Na$_2$O~+~K$_2$O. Composition data for samples that did not exhibit a zero-field ESR signal are given in Table~S1.}
\label{tab:samples}
\end{table*}

Clock transitions (CTs) in spin systems can further enhance quantum coherence by protecting the qubit from the environmental spin bath~\cite{takahashiDecoherenceCrystalsQuantum2011,wolfowiczAtomicClockTransitions2013,zadroznyMillisecondCoherenceTime2015,shiddiqEnhancingCoherenceMolecular2016,collettClockTransitionCr7Mn2019,kundu2GHzClockTransition2022} and can be used for both single-qubit and multi-qubit gates~\cite{collettConstructingClocktransitionbasedTwoqubit2020,littleExperimentalRealisationMultiqubit2023}. A CT occurs at an avoided level crossing where $\nabla_\mathbf{B} \omega = \mathbf{0}$; the level transition frequency $\omega$ is therefore immune to the decohering effects of small magnetic field fluctuations to first order. One of the attractive features of MNMs is their chemical engineerability -- their properties, including the existence  and ESR frequency of a CT in the MNM~\cite{laorenzaTunableCrMolecular2021,baylissEnhancingSpinCoherence2022}, can be designed and controlled through synthetic processes. In contrast, ``natural'' CTs come ready made and have the potential to be more robust. Here we present a study of previously undiscovered zero-field CTs in certain types of defect-rich structural silica glasses. In these CTs, we find coherence times over 15~$\upmu$s, a value that makes these systems attractive as potential qubits. While the underlying physics of these CTs is not yet definitively determined, we present evidence to support the hypothesis that the CTs arise from spin-1 boron-vacancy or aluminum-vacancy centers within the silica matrix. With future work, these silica glasses could be engineered to provide further-enhanced coherence at their CTs.

The usefulness of these glass spin systems within hybrid architectures provides further reason for study. While the addressing of individual qubits within a glass may be challenging -- unless they are also color centers that can be addressed optically~\cite{billaudMicrowaveFluorescenceDetection2023,mullinQuantumSensingMagnetic2023} -- the silica CTs, and their enhanced coherence times, may be useful as a platform for quantum memory in which information is stored holographically within the entire material~\cite{mortonStoringQuantumInformation2018}. As such, a spin system with a CT can be integrated into a hybrid architecture wherein the spin qubit is strongly coupled to other qubits~\cite{kuboHybridQuantumCircuit2011}. A zero-field CT permits coupling to superconducting qubit systems, which are sensitive to the applied magnetic fields typically used for spin qubits. Here, there is no requirement of applying a magnetic field to address the spin states of the zero-field CT spin qubit, which can be coupled to superconducting qubits without this additional complication.

Silica glass is the glass form of quartz, made of SiO$_2$ in a random matrix. The magnetic properties of defects in quartz and other forms of silica have been studied over the years for a number of different purposes, including: thermal~\cite{stephensIntrinsicLowtemperatureThermal1976}, optical~\cite{griscomOpticalPropertiesStructure1991}, and magnetic~\cite{jugGlassyStateMagnetically2014,jugRevealingIntrinsicMagnetism2021} properties, and generation of spin-1/2 E$^\prime$ centers via implantation and irradiation~\cite{weeksManyVarietiesCenters1994,magruderEffectImplantingBoron2008,wangGIrradiationEffectsBorosilicate2020}. Spin systems in quartz after irradiation, however, tend to be spin-1/2 systems with a Zeeman-like Hamiltonian and $g\sim2$. Furthermore, many borosilicate glasses and quartzes have iron (Fe$^{3+}$) impurities, which are interstitial defects with an electronic $S=5/2$ and $g\approx4.3$. Silica glasses have previously been shown to have vacancies, as well, including charged oxygen vacancies~\cite{kimmelDoublyPositivelyCharged2009,kimmelPositiveNegativeOxygen2009}. B$_2$O$_3$ in glass and crystalline phases has been examined at X-band after irradiation, showing significant CW ESR signal near $g=2$ and electron-spin echo envelope modulation (ESEEM) via pulsed ESR associated with oxygen dangling bonds in boron trigonal units~\cite{deligiannakisElectronicStructureGlass1998}. Still other glasses, including several of the glasses examined for the current study, have been examined for zero-field increase in polarization echoes, which were not used to quantify spin coherence~\cite{ludwigMagneticFieldDependent2003,nagelNovelIsotopeEffects2004}. The potential of these systems for use as spin qubits, therefore, has never been examined.  One attractive feature of silica for quantum coherence is the fact that the majority isotopes of silicon (92.2\% $^{28}$Si, 3.1\% $^{30}$Si) and oxygen (99.8\% $^{16}$O, 0.2\% $^{18}$O) have zero nuclear spin, so the  glass matrix intrinsically has very low nuclear magnetic noise.

Although the energy-level structure of the CT spin system in silica is not yet fully characterized, we provide a heuristic model of an avoided crossing at zero field. Generation of an avoided crossing at zero field requires a system with integer spin due to Kramers' Theorem; here, we employ a working model of a spin-1 system with spin Hamiltonian
\begin{equation}
 \label{eqn:spin1ham}
 \mathcal{H}=-D S_z^2+E (S_x^2-S_y^2)+g_s \mu_B\boldsymbol{B}\cdot\boldsymbol{S}.
\end{equation}
The first and second terms represent longitudinal and transverse anisotropies, respectively, and the third term is the Zeeman interaction with the magnetic field. In the $\hat{S}_z$ basis, the energy eigenstates at zero field are 
\begin{equation}
\begin{aligned}
 &\ket{+} =(\ket{1}+\ket{-1})/\sqrt{2},\quad \\
 &\ket{-}=(\ket{1}-\ket{-1})/\sqrt{2},\\
 &\ket{0},\text{} 
\end{aligned}
\end{equation}
with energies $-D+E,-D-E$, and 0, respectively. With $D\gg E$, the avoided crossing becomes an effective two-level system. The energy splitting of the avoided crossing is therefore $2E$, and ESR radiation applied such that $\hbar\omega = 2E$ addresses the CT.

\begin{figure}[ht!]
 \centering
 \includegraphics[width=1\linewidth]{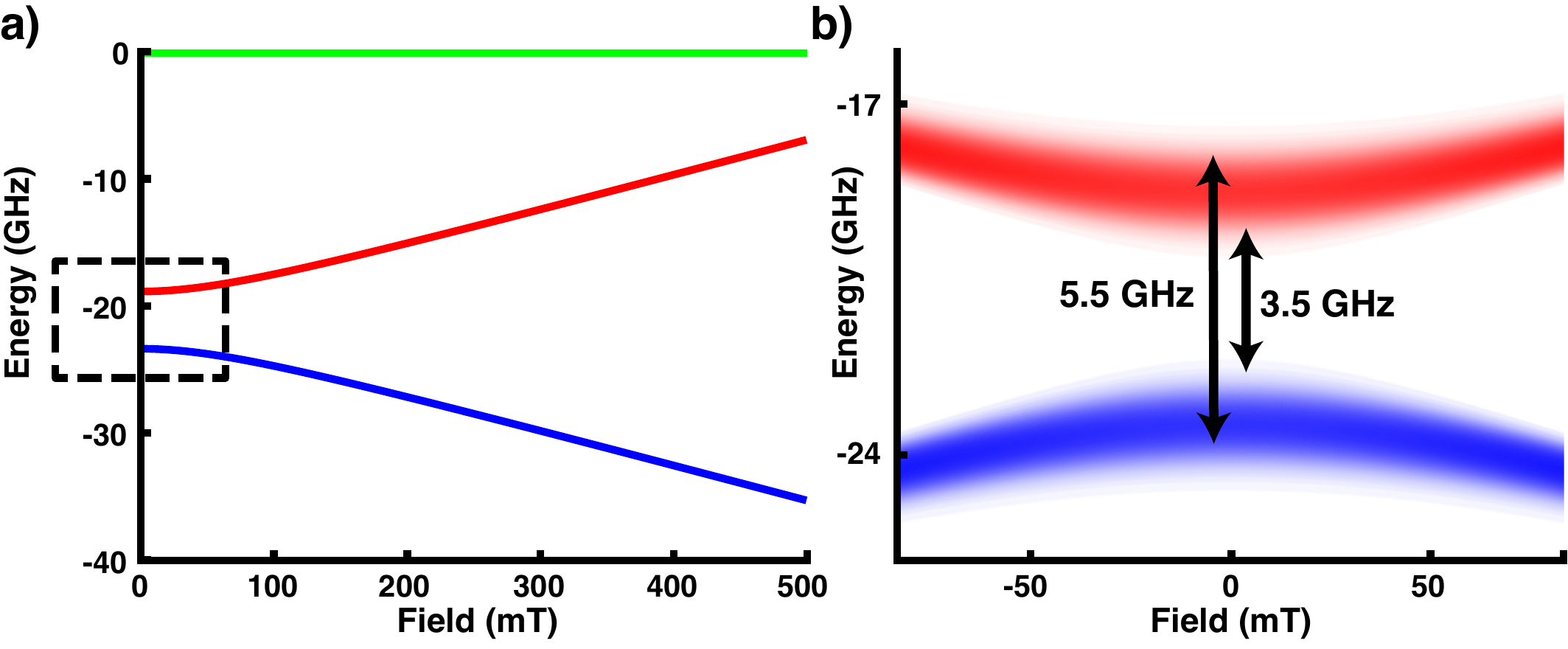}
 \caption{\textbf{Level diagrams of a spin-1 system.} (a) An energy level diagram of a spin-1 system. Green represents the $\ket{0}$ state, while red and blue, the two lowest-lying states, mix at zero field to form an avoided crossing. The level diagram was calculated using a spin Hamiltonian shown in Eq.~\ref{eqn:spin1ham}, with $D=21$~GHz and $E=2.25$~GHz. The dashed box around the avoided crossing of the blue and red levels shows the inset for panel (b). (b) A sketch of an inhomogeneously broadened energy level diagram depicting the avoided crossing at zero field. This inhomogeneous broadening was drawn with a Gaussian distribution in $E$ with a FWHM of 1.0~GHz. Inhomogeneous broadening in $E$ provides a mechanism of generating a broadband CT, matching the behavior seen in the spin systems shown in this study. The arrows represent the edges of the frequency range over which we measure the clock transition.}
 \label{fig:spinHam_inhomo}
\end{figure}

Fig.~\ref{fig:spinHam_inhomo}(a) shows the energy-level diagram calculated from the eigenstates of Eq.~\ref{eqn:spin1ham} with $D=21$~GHz and $E=2.25$~GHz and assuming the $\Vec{B}\parallel\hat{z}$. The states drawn in red and blue show the states that mix to form the $\ket{+}$ and $\ket{-}$ states, which constitute the two-level system. In our experiments, we observe a substantial range of ESR frequencies that give rise to CT behavior at zero field (Fig.~S4 of the SI provides examples of the broad frequency accessibility of the CT). Indeed, we observe signal over a wide range of frequencies from $\sim$3.5~GHz to 5.5~GHz. Within this range, the signal amplitude and spin dynamics at the CT is largely unchanged. This indicates significant inhomogeneity in the transverse anisotropy such that at any given ESR frequency only a small subensemble of the total spins in the sample is addressed. To account for this observation, we include a Gaussian distribution in $E$ into our heuristic model. Fig.~\ref{fig:spinHam_inhomo}(b) (corresponding to the dashed black box in (a)) shows the avoided crossing generated using Eq.~\ref{eqn:spin1ham} with $D=21$~GHz and a Gaussian distribution in $E$ centered on $E = 2.25$~GHz with a FWHM of $1.0$~GHz. The intensity of the color for both states corresponds to the density of spins at each $E$-value in the distribution. The black arrows indicate the range of frequencies over which the CT is observed to be strongest.

\section{Methods}

To perform ESR experiments, all samples were mounted into a loop-gap resonator (LGR)~\cite{fronciszLoopgapResonatorNew1982,eisenachBroadbandLoopGap2018} made of OFHC copper and held in place with a small quantity of vacuum grease. Measurements were done using a home-built ESR probe with \textit{in situ} frequency and coupling tunability~\cite{joshiAdjustableCouplingSitu2020} within a Quantum Design Physical Property Measurement System (PPMS) with a base temperature of 1.8~K. Typically, our resonators have a high quality factor ($Q\sim2000$ at low temperature); for pulse ESR experiments, a small quantity of Eccosorb-brand microwave absorber was placed into the sample chamber near the resonator to reduce the $Q$ and reduce ringdown times. For all experiments reported herein, the LGR was mounted into the probe such that the microwave $B_1$ field of the resonator was parallel to the DC field $B_0$. The spectrometer was built from commercially available microwave circuit components; microwave pulses were generated via the programmable Tabor SE5082 Arbitrary Waveform Generator (AWG) and echoes were measured via a homodyne technique and digitized using an oscilloscope. To remove background signals, we employed a four-step phase-cycling process.

Table~\ref{tab:samples} provides chemical compositions for  glass samples examined in this study. Duran is a brand name for borosilicate glass 3.3 (DIN ISO 3585); aluminosilicate 0812 is manufactured by Schott; Vycor 7913 was manufactured by Corning; and AGC40 is manufactured by Advanced Glass and Ceramics to mimic the physical properties of Vycor 7913. Both Vycor 7913 and AGC40 are porous borosilicate glass.  One sample, that we simply call ``borosilicate'' glass was sourced from McMaster-Carr, which is a distributor, not a manufacturer. The provenance and detailed chemical characterization of that sample is not established.

\begin{figure}[ht!]
 \centering
 \includegraphics[width=1\linewidth]{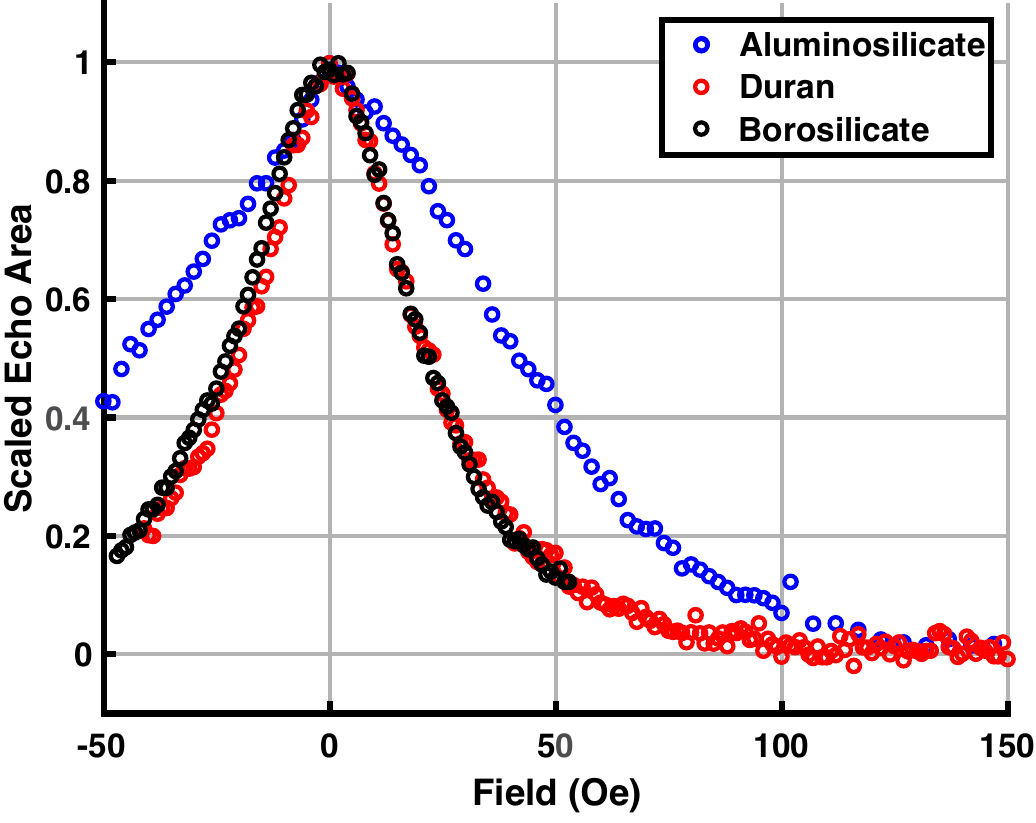}
 \caption{\textbf{Echo signal dependence on magnetic field.} Echo response with a standard Hahn echo sequence near zero field of aluminosilicate glass (measured at 5036~MHz, delay time $\uptau = 1000$~ns, blue), Duran glass (3975~MHz, $\uptau = 800$~ns, red), and borosilicate glass (5117~MHz, $\uptau = 1000$~ns, black). All echo data was measured at 1.8~K and within the frequency range of the inhomogeneously broadened clock transition (Fig.~\ref{fig:spinHam_inhomo}).}
 \label{fig:EDFS_CT}
\end{figure}

\section{Results}

\begin{figure*}[ht!]
 \centering
 \includegraphics[width=1\linewidth]{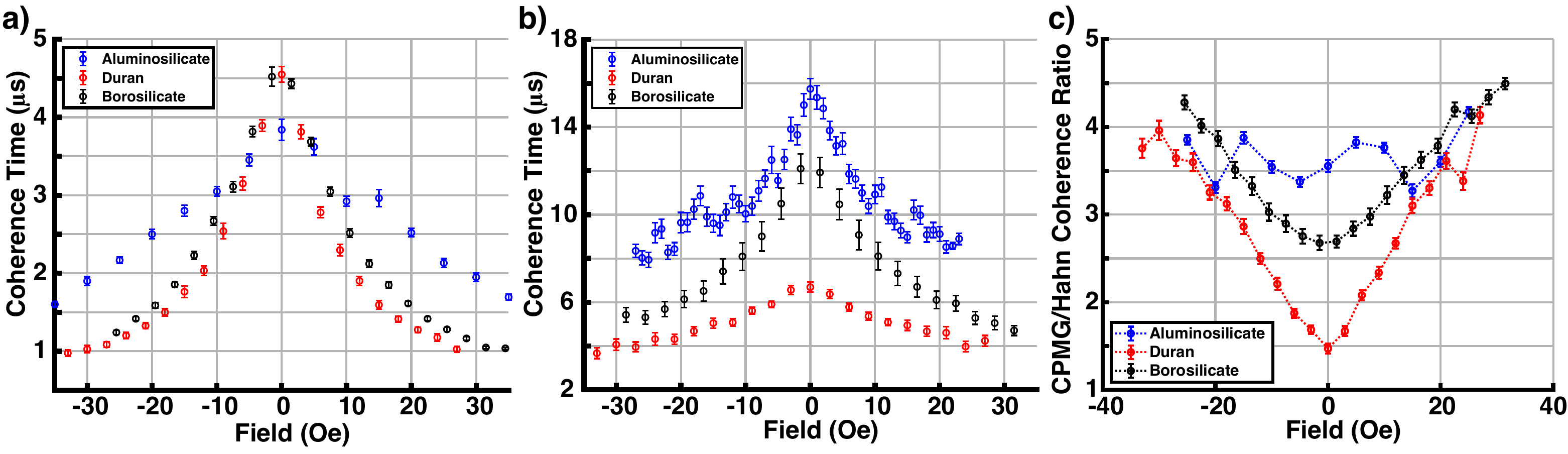}
 \caption{\textbf{Coherence at zero field.} (a) Shows the increase in coherence time, as measured via a Hahn echo sequence, as field is tuned to zero. Aluminosilicate (blue), Duran (red), and borosilicate (black) glasses all show a peak at zero field. When paired with the data in Fig.~\ref{fig:EDFS_CT}, these data have the hallmarks of a zero-field clock transition: enhanced echo signal and coherence. A Vycor 7913 sample also showed a zero-field-peaked ESR signal, albeit much weaker than that found in other glass samples that showed CTs (Fig.~S3).  (b) $T_2$ measured via the CPMG pulse sequence for the same three samples (with delay $\uptau = 1000$~ns for aluminosilicate and borosilicate, and $\uptau = 800$~ns for Duran). (c) The ratio of $T_2$ measured by CPMG to $T_2$ measured by Hahn echo, for all three glass samples. For each panel, all experiments were performed at 1.8~K, with ESR frequencies of 5036~MHz (aluminosilicate), 4028~MHz (Duran), and 5117~MHz (borosilicate).}
 \label{fig:coherence_CT}
\end{figure*}

Fig.~\ref{fig:EDFS_CT} shows the results of studying Duran glass (13\% B$_2$O$_3$-in-SiO$_2$), red, a stoichiometrically similar borosilicate glass sample, black, and a sample of 0812 aluminosilicate glass, blue, using pulsed ESR as a function of magnetic field. A standard Hahn echo sequence was used. All three samples show a clearly defined zero-field peak, although the aluminosilicate's peak is wider with a cusp at zero field while the Duran's and borosilicate's peaks are more rounded. While the data for each sample shown in  Fig.~\ref{fig:EDFS_CT} was taken at  a different microwave frequency, we have not noted any appreciable difference in sample behavior within a wide range of frequencies -- see discussion of  Fig.~\ref{fig:spinHam_inhomo}(b) above. Fig.~S4 of the SI shows examples of the CT being addressed across the range $\sim3.5$~GHz -- $\sim5.1$~GHz; the CT demonstrates qualitatively similar behavior throughout. 

To further characterize these signals, we study the coherence of these transitions by examining the dependence of the echo area on the delay time $\uptau$ between the $\uppi/2$ and $\uppi$ pulses. An example of this dependence is shown in Fig.~S1(a). Fig.~\ref{fig:coherence_CT}(a) shows $T_2$ for aluminosilicate (blue), Duran (red), and borosilicate (black) glasses. In each sample $T_2$ is longest (up to $\sim4$~$\upmu$s) at zero field, and falls off rapidly as the field is tuned away from zero. This observation is the hallmark of the CT effect: To first order, spins in the system are protected from the decohering effects of dipolar interactions, and can therefore retain their phase information for longer.  We have also observed Rabi oscillations at the CT in our samples (Fig.~S2).

By applying the Carr-Purcell-Meiboom-Gill (CPMG) pulse sequence, wherein repeated application of $\uppi$ pulses refocuses the spins via the sequence $(\uppi/2)_x - (\tau - \uppi_y - \tau - \text{echo})^N$ (for $N$ integer), coherence can be further enhanced. The decay of the echo area as a function of total evolution time provides a measure of this enhanced $T_2$. Fig.~\ref{fig:coherence_CT}(b) shows that each sample exhibits enhanced coherence with the CPMG pulse sequence (when compared with the Hahn results of panel (a)). An example of a CPMG trace is shown in Fig.~S1(b). While all three samples see a clear CPMG enhancement of $T_2$, the effect is material dependent. Both borosilicate samples show similar $T_2$ from the Hahn experiment (panel a), but the Duran shows a much smaller CPMG enhancement in panel (b). The aluminosilicate shows the largest enhancement, with coherence times measured by CPMG exceeding 15~$\upmu$s. We note that the aluminosilicate sample also shows a series of small side peaks, seen most clearly in Fig.~\ref{fig:coherence_CT}(b) between 10~Oe and 20~Oe, -- perhaps arising from hyperfine coupling to the $^{27}$Al nuclear spin -- but these were not investigated in detail in this study. 

Fig.~\ref{fig:coherence_CT}(c) shows the ratio ($T_{2,\text{ CPMG}}/T_{2,\text{ Hahn echo}}$) of coherence times measured in Figs.~\ref{fig:coherence_CT}(a,b) for all three glass samples as a function of magnetic field. Each sample behaves differently: though borosilicate (black) and Duran (red) both show that CPMG is less effective at enhancing coherence at the CT than when the field is tuned away, Duran shows that CPMG only improves coherence by a factor of 1.5 at the CT. One interpretation is that as this ratio approaches unity at the CT, CPMG becomes redundant as a decoherence noise filter since the CT already filters low-frequency magnetic noise. This suggests that the primary source of decoherence in this material is magnetic noise. Aluminosilicate (blue) shows no such relationship -- no clear trend in the ratio of coherence times can be seen as the field is tuned through the CT.

\begin{figure}[ht!]
 \centering
 \includegraphics[width=1\linewidth]{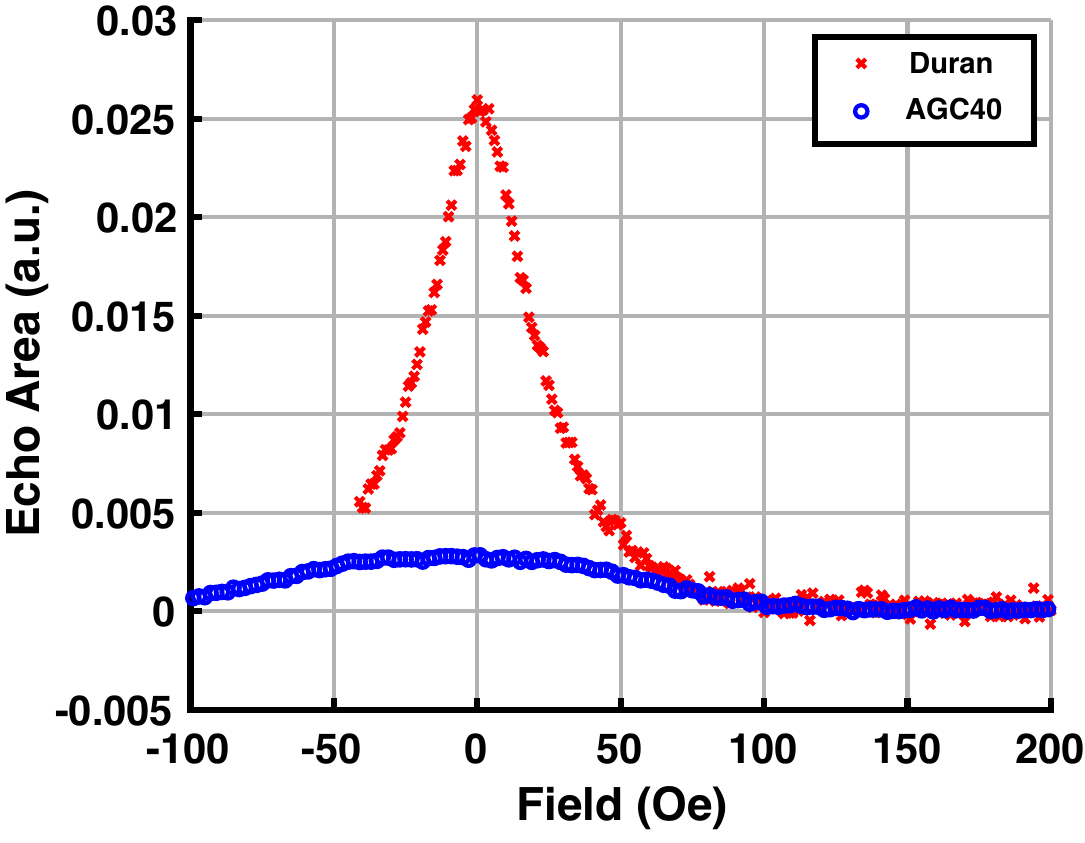}
 \caption{\textbf{Comparison of EDFS at CT with different concentrations of boron in borosilicate glasses.} Echo area, from a standard Hahn sequence with delay $\uptau = 1000$~ns, as a function of magnetic field near the clock transition for two different samples of borosilicate glass, showing different intensity of the peak at the clock transition. In red, the clock transition for Duran glass is shown (measured at 1.8~K and 3975~MHz); in blue, the clock transition for the AGC40 glass (measured at 1.8~K and 3773~MHz) shows much smaller echo signal at zero field.}
 \label{fig:concentrationcomparison}
\end{figure}

One commonality among the samples exhibiting the CT effect is the presence of both SiO$_2$ and boron as a substitutional defect. Thus, we examined multiple borosilicate glasses with differing boron concentrations. Fig.~\ref{fig:concentrationcomparison} shows a comparison of 4\% B$_2$O$_3$ in SiO$_2$ (in blue – AGC40) and Duran glass (also shown in Fig.~\ref{fig:EDFS_CT} in red), overlaid. At zero field the echo area is substantially reduced in the dilute sample, suggesting that the spin system producing the CT has been reduced. The ratio of signal sizes is not 13:4, as one might initially expect. Part of this may be due to small changes in experimental conditions that change absolute signal size between experimental runs. Nevertheless, there are clearly other differences between the two samples with the peak for AGC40 being wider and having a much flatter top than for Duran. A confounding factor is that AGC40 is porous while Duran is not, making comparisons between the two samples challenging.

\begin{figure}[ht!]
 \centering
 \includegraphics[width=1\linewidth]{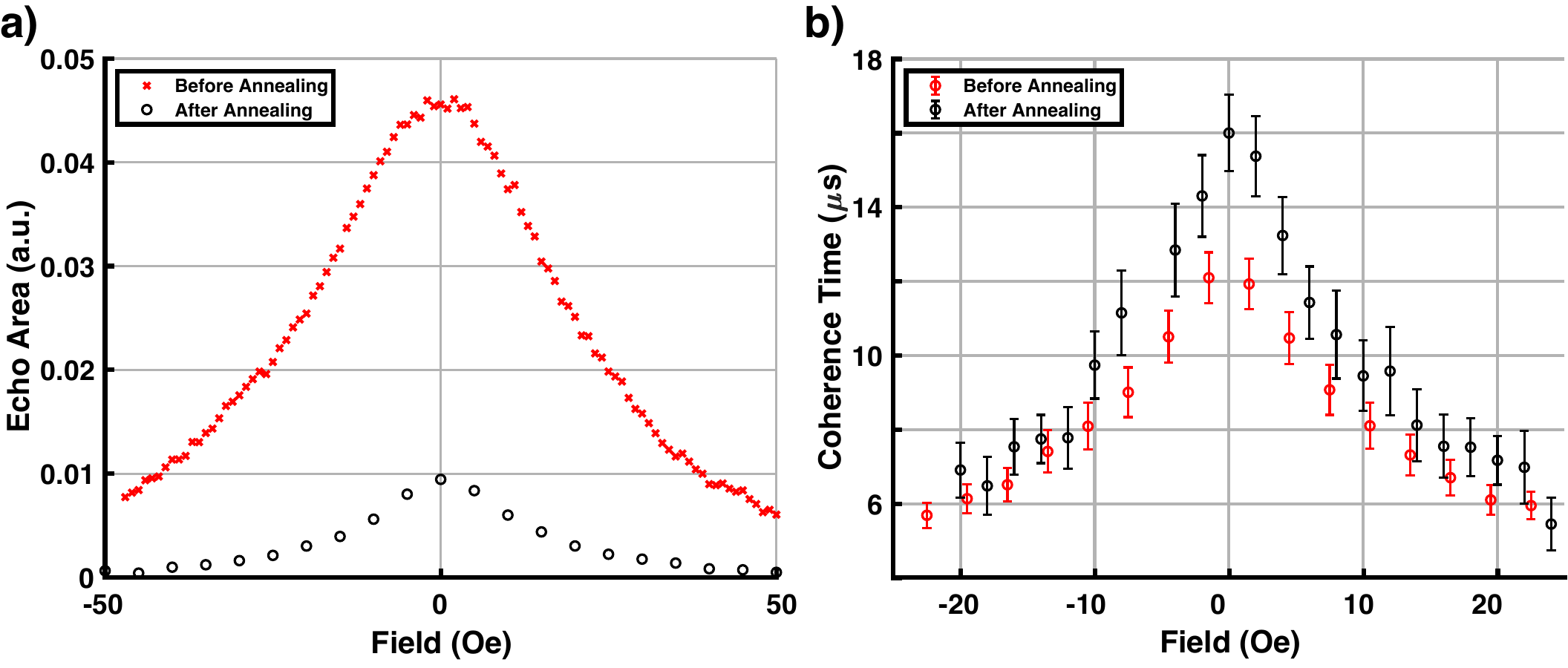}
 \caption{\textbf{Comparison of CT before and after annealing borosilicate glass.} (a) A comparison of the echo area at the CT before (red, measured at 1.8~K and 5117~MHz) and after (black, measured at 1.8~K and 3836~MHz) annealing the sample. After annealing the sample, echo signal drops substantially at and away from the CT, suggesting that fewer spins in the annealed sample are being addressed by the working ESR frequency. (b) Overlay of coherence times of the same samples as (a), measured by the CPMG pulse sequence, showing a $\sim$32\% increase in $T_2$ at zero field after annealing the 13\% borosilicate glass. All data in both panels used a delay time of $\uptau = 1000$~ns. Table~S2 in the Supplementary Information provides details of the annealing process for all samples. We kept the ESR frequency for all data shown restricted to within the frequency range of the inhomogeneously broadened CT.}
 \label{fig:boroanneal}
\end{figure}

We also examined the behavior of each sample after annealing. Details of the annealing process for each annealed sample are provided in Table~S2 of the Supplementary Information. AGC40, Vycor 7913, and aluminosilicate glass showed no echo after annealing. Figure~\ref{fig:boroanneal}(a) shows a drastically reduced echo signal after annealing for the borosilicate; the echo signal at zero field is reduced by nearly 80\%. This suggests that annealing the glass reduces the number of the spin centers generating the CT. Fig.~\ref{fig:boroanneal}(b) shows a remarkable $T_2$ value of $\sim$16~$\upmu$s at the CT as measured with CPMG, a 32\% increase over the non-annealed sample. While annealing substantially reduced the CT signal for nearly all of our samples, one notable exception was the Duran sample, which showed no significant change. This distinction will be the subject of further investigation.

\section{Discussion}

Our results demonstrate substantial coherence times in several glass samples, with  $T_2$ enhanced at the zero-field CT.  ESR response for our CT signal also appears to depend on boron concentration (e.g.~Fig.~\ref{fig:concentrationcomparison}).  In addition, annealing  reduces the response while simultaneously enhancing coherence. While annealing may have many effects on the properties of the glass, one well-known consequence of the annealing process is the removal of vacancies. Silica glasses and quartz are known to contain numerous vacancies, notably oxygen vacancies~\cite{skujaDefectsOxideGlasses2005,kimmelDoublyPositivelyCharged2009}. Our observations then suggest that vacancies potentially play a role in the spin system that is responsible for the CT. One hypothesis, then, for  the microscopic source of the CT in the borosilicate glasses is a spin-1 boron vacancy center (and, for aluminosilicate glass, a spin-1 aluminum vacancy center). Boron and aluminum are known substitutional vacancies for silicon, widely used as dopants in semiconductors. Both in Group 13 of the periodic table, they are electron acceptors when substituting for silicon. They therefore introduce a localized hole with spin 1/2 when doped. We hypothesize that when coupled to a neighboring charged oxygen vacancy that also has a spin 1/2, such defects can form a spin-1 system capable of producing a zero-field CT. One tantalizing, but perhaps remote, possibility is that this center may behave as a molecular color center in silica glass, a feature with obvious applications in quantum sensing, analogous to the well-known NV$^-$ center in diamond, or more specifically the negatively charged BV center in diamond~\cite{muruganathanBoronVacancyColor2021,umedaNegativelyChargedBoron2022}.

Regardless of the microscopic origin of the CT, the long values of $T_2$ observed here are remarkable.  The materials are ordinary, commercially available glasses that have not been engineered for quantum information applications.  Thus, with the development of a detailed understanding of the defect underlying the CT and careful engineering, much larger coherence times could be on the horizon.  We hope that through continued experimental work, materials engineering and theoretical analysis, we may uncover a deeper understanding of the nature of the underlying CT, and develop methods to further enhance $T_2$.

\section{Author Contributions}
The experiments were conceived of by B.C.S, G.C., and J.R.F., and carried out by B.C.S and G.C. Sample annealing was performed by B.C.S. B.C.S. performed the data analysis. B.C.S. and J.R.F. wrote the manuscript, with input from G.C.

The authors declare no competing financial interest.

\section{Acknowledgements}

We thank G.~Joshi and K.-I.~Ellers for carrying out some preliminary experiments, J.~Kubasek for assistance with machining the resonators used in this study, and B.~Crepeau for assistance with instrumentation. C.~A.~Collett, I.~Nikolov, F.~Abdo~Arias, T.~Pathak, K.~Thompson, and J.~Levy provided useful conversations and advice. We are grateful to S.~Prasch for help with annealing samples. This work was supported by the Research Corporation for Science Advancement under Cottrell SEED Award No.~27849, and by the National Science Foundation under grant nos.~DMR-1708692 and DMR-2207624. JRF acknowledges the support of the Amherst College Senior Sabbatical Fellowship Program, which is funded in part by the H.~Axel Schupf~’57 Fund for Intellectual Life.

\bibliography{Silica_Defects_and_Clock_Transitions}

\providecommand{\latin}[1]{#1}
\makeatletter
\providecommand{\doi}
  {\begingroup\let\do\@makeother\dospecials
  \catcode`\{=1 \catcode`\}=2 \doi@aux}
\providecommand{\doi@aux}[1]{\endgroup\texttt{#1}}
\makeatother
\providecommand*\mcitethebibliography{\thebibliography}
\csname @ifundefined\endcsname{endmcitethebibliography}  {\let\endmcitethebibliography\endthebibliography}{}
\begin{mcitethebibliography}{47}
\providecommand*\natexlab[1]{#1}
\providecommand*\mciteSetBstSublistMode[1]{}
\providecommand*\mciteSetBstMaxWidthForm[2]{}
\providecommand*\mciteBstWouldAddEndPuncttrue
  {\def\EndOfBibitem{\unskip.}}
\providecommand*\mciteBstWouldAddEndPunctfalse
  {\let\EndOfBibitem\relax}
\providecommand*\mciteSetBstMidEndSepPunct[3]{}
\providecommand*\mciteSetBstSublistLabelBeginEnd[3]{}
\providecommand*\EndOfBibitem{}
\mciteSetBstSublistMode{f}
\mciteSetBstMaxWidthForm{subitem}{(\alph{mcitesubitemcount})}
\mciteSetBstSublistLabelBeginEnd
  {\mcitemaxwidthsubitemform\space}
  {\relax}
  {\relax}

\bibitem[Kjaergaard \latin{et~al.}(2020)Kjaergaard, Schwartz, Braum{\"u}ller, Krantz, Wang, Gustavsson, and Oliver]{kjaergaardSuperconductingQubitsCurrent2020}
Kjaergaard,~M.; Schwartz,~M.~E.; Braum{\"u}ller,~J.; Krantz,~P.; Wang,~J. I.-J.; Gustavsson,~S.; Oliver,~W.~D. Superconducting {{Qubits}}: {{Current State}} of {{Play}}. \emph{Annu. Rev. Condens. Matter Phys.} \textbf{2020}, \emph{11}, 369--395\relax
\mciteBstWouldAddEndPuncttrue
\mciteSetBstMidEndSepPunct{\mcitedefaultmidpunct}
{\mcitedefaultendpunct}{\mcitedefaultseppunct}\relax
\EndOfBibitem
\bibitem[Bruzewicz \latin{et~al.}(2019)Bruzewicz, Chiaverini, McConnell, and Sage]{bruzewiczTrappedionQuantumComputing2019}
Bruzewicz,~C.~D.; Chiaverini,~J.; McConnell,~R.; Sage,~J.~M. Trapped-Ion Quantum Computing: {{Progress}} and Challenges. \emph{Appl. Phys. Rev.} \textbf{2019}, \emph{6}, 021314\relax
\mciteBstWouldAddEndPuncttrue
\mciteSetBstMidEndSepPunct{\mcitedefaultmidpunct}
{\mcitedefaultendpunct}{\mcitedefaultseppunct}\relax
\EndOfBibitem
\bibitem[Morton \latin{et~al.}(2011)Morton, McCamey, Eriksson, and Lyon]{mortonEmbracingQuantumLimit2011}
Morton,~J. J.~L.; McCamey,~D.~R.; Eriksson,~M.~A.; Lyon,~S.~A. Embracing the Quantum Limit in Silicon Computing. \emph{Nature} \textbf{2011}, \emph{479}, 345--353\relax
\mciteBstWouldAddEndPuncttrue
\mciteSetBstMidEndSepPunct{\mcitedefaultmidpunct}
{\mcitedefaultendpunct}{\mcitedefaultseppunct}\relax
\EndOfBibitem
\bibitem[Heinrich \latin{et~al.}(2021)Heinrich, Oliver, Vandersypen, Ardavan, Sessoli, Loss, Jayich, {Fernandez-Rossier}, Laucht, and Morello]{heinrichQuantumcoherentNanoscience2021}
Heinrich,~A.~J.; Oliver,~W.~D.; Vandersypen,~L. M.~K.; Ardavan,~A.; Sessoli,~R.; Loss,~D.; Jayich,~A.~B.; {Fernandez-Rossier},~J.; Laucht,~A.; Morello,~A. Quantum-Coherent Nanoscience. \emph{Nat. Nanotechnol.} \textbf{2021}, \emph{16}, 1318--1329\relax
\mciteBstWouldAddEndPuncttrue
\mciteSetBstMidEndSepPunct{\mcitedefaultmidpunct}
{\mcitedefaultendpunct}{\mcitedefaultseppunct}\relax
\EndOfBibitem
\bibitem[Coronado(2019)]{coronadoMolecularMagnetismChemical2019}
Coronado,~E. Molecular Magnetism: From Chemical Design to Spin Control in Molecules, Materials and Devices. \emph{Nat. Rev. Mater.} \textbf{2019}, \emph{5}, 87--104\relax
\mciteBstWouldAddEndPuncttrue
\mciteSetBstMidEndSepPunct{\mcitedefaultmidpunct}
{\mcitedefaultendpunct}{\mcitedefaultseppunct}\relax
\EndOfBibitem
\bibitem[Chiesa \latin{et~al.}(2024)Chiesa, Santini, Garlatti, Luis, and Carretta]{chiesaMolecularNanomagnetsViable2024}
Chiesa,~A.; Santini,~P.; Garlatti,~E.; Luis,~F.; Carretta,~S. Molecular Nanomagnets: A Viable Path toward Quantum Information Processing? \emph{Rep. Prog. Phys.} \textbf{2024}, \emph{87}, 034501\relax
\mciteBstWouldAddEndPuncttrue
\mciteSetBstMidEndSepPunct{\mcitedefaultmidpunct}
{\mcitedefaultendpunct}{\mcitedefaultseppunct}\relax
\EndOfBibitem
\bibitem[Rabl \latin{et~al.}(2006)Rabl, DeMille, Doyle, Lukin, Schoelkopf, and Zoller]{rablHybridQuantumProcessors2006}
Rabl,~P.; DeMille,~D.; Doyle,~J.~M.; Lukin,~M.~D.; Schoelkopf,~R.~J.; Zoller,~P. Hybrid {{Quantum Processors}}: {{Molecular Ensembles}} as {{Quantum Memory}} for {{Solid State Circuits}}. \emph{Phys. Rev. Lett.} \textbf{2006}, \emph{97}, 033003\relax
\mciteBstWouldAddEndPuncttrue
\mciteSetBstMidEndSepPunct{\mcitedefaultmidpunct}
{\mcitedefaultendpunct}{\mcitedefaultseppunct}\relax
\EndOfBibitem
\bibitem[Morton and Lovett(2011)Morton, and Lovett]{mortonHybridSolidStateQubits2011}
Morton,~J.~J.; Lovett,~B.~W. Hybrid {{Solid-State Qubits}}: {{The Powerful Role}} of {{Electron Spins}}. \emph{Annu. Rev. Condens. Matter Phys.} \textbf{2011}, \emph{2}, 189--212\relax
\mciteBstWouldAddEndPuncttrue
\mciteSetBstMidEndSepPunct{\mcitedefaultmidpunct}
{\mcitedefaultendpunct}{\mcitedefaultseppunct}\relax
\EndOfBibitem
\bibitem[Clerk \latin{et~al.}(2020)Clerk, Lehnert, Bertet, Petta, and Nakamura]{clerkHybridQuantumSystems2020}
Clerk,~A.~A.; Lehnert,~K.~W.; Bertet,~P.; Petta,~J.~R.; Nakamura,~Y. Hybrid Quantum Systems with Circuit Quantum Electrodynamics. \emph{Nat. Phys.} \textbf{2020}, \emph{16}, 257--267\relax
\mciteBstWouldAddEndPuncttrue
\mciteSetBstMidEndSepPunct{\mcitedefaultmidpunct}
{\mcitedefaultendpunct}{\mcitedefaultseppunct}\relax
\EndOfBibitem
\bibitem[Gimeno \latin{et~al.}(2020)Gimeno, Kersten, Pallar{\'e}s, Hermosilla, {Mart{\'i}nez-P{\'e}rez}, Jenkins, Angerer, {S{\'a}nchez-Azqueta}, Zueco, Majer, Lostao, and Luis]{gimenoEnhancedMolecularSpinPhoton2020}
Gimeno,~I.; Kersten,~W.; Pallar{\'e}s,~M.~C.; Hermosilla,~P.; {Mart{\'i}nez-P{\'e}rez},~M.~J.; Jenkins,~M.~D.; Angerer,~A.; {S{\'a}nchez-Azqueta},~C.; Zueco,~D.; Majer,~J.; Lostao,~A.; Luis,~F. Enhanced {{Molecular Spin-Photon Coupling}} at {{Superconducting Nanoconstrictions}}. \emph{ACS Nano} \textbf{2020}, acsnano.0c03167\relax
\mciteBstWouldAddEndPuncttrue
\mciteSetBstMidEndSepPunct{\mcitedefaultmidpunct}
{\mcitedefaultendpunct}{\mcitedefaultseppunct}\relax
\EndOfBibitem
\bibitem[Luis \latin{et~al.}(2000)Luis, Mettes, Tejada, Gatteschi, and De~Jongh]{luisObservationQuantumCoherence2000}
Luis,~F.; Mettes,~F.~L.; Tejada,~J.; Gatteschi,~D.; De~Jongh,~L.~J. Observation of {{Quantum Coherence}} in {{Mesoscopic Molecular Magnets}}. \emph{Phys. Rev. Lett.} \textbf{2000}, \emph{85}, 4377--4380\relax
\mciteBstWouldAddEndPuncttrue
\mciteSetBstMidEndSepPunct{\mcitedefaultmidpunct}
{\mcitedefaultendpunct}{\mcitedefaultseppunct}\relax
\EndOfBibitem
\bibitem[Leuenberger and Loss(2001)Leuenberger, and Loss]{leuenbergerQuantumComputingMolecular2001}
Leuenberger,~M.~N.; Loss,~D. Quantum Computing in Molecular Magnets. \emph{Nature} \textbf{2001}, \emph{410}, 789--793\relax
\mciteBstWouldAddEndPuncttrue
\mciteSetBstMidEndSepPunct{\mcitedefaultmidpunct}
{\mcitedefaultendpunct}{\mcitedefaultseppunct}\relax
\EndOfBibitem
\bibitem[Tejada \latin{et~al.}(2001)Tejada, Chudnovsky, del Barco, Hernandez, and Spiller]{tejadaMagneticQubitsHardware2001}
Tejada,~J.; Chudnovsky,~E.~M.; del Barco,~E.; Hernandez,~J.~M.; Spiller,~T.~P. Magnetic Qubits as Hardware for Quantum Computers. \emph{Nanotechnology} \textbf{2001}, \emph{12}, 181--186\relax
\mciteBstWouldAddEndPuncttrue
\mciteSetBstMidEndSepPunct{\mcitedefaultmidpunct}
{\mcitedefaultendpunct}{\mcitedefaultseppunct}\relax
\EndOfBibitem
\bibitem[Friedman and Sarachik(2010)Friedman, and Sarachik]{friedmanSingleMoleculeNanomagnets2010}
Friedman,~J.~R.; Sarachik,~M.~P. Single-{{Molecule Nanomagnets}}. \emph{Annu. Rev. Condens. Matter Phys.} \textbf{2010}, \emph{1}, 109--128\relax
\mciteBstWouldAddEndPuncttrue
\mciteSetBstMidEndSepPunct{\mcitedefaultmidpunct}
{\mcitedefaultendpunct}{\mcitedefaultseppunct}\relax
\EndOfBibitem
\bibitem[Takahashi \latin{et~al.}(2011)Takahashi, Tupitsyn, Van~Tol, Beedle, Hendrickson, and Stamp]{takahashiDecoherenceCrystalsQuantum2011}
Takahashi,~S.; Tupitsyn,~I.~S.; Van~Tol,~J.; Beedle,~C.~C.; Hendrickson,~D.~N.; Stamp,~P. C.~E. Decoherence in Crystals of Quantum Molecular Magnets. \emph{Nature} \textbf{2011}, \emph{476}, 76--79\relax
\mciteBstWouldAddEndPuncttrue
\mciteSetBstMidEndSepPunct{\mcitedefaultmidpunct}
{\mcitedefaultendpunct}{\mcitedefaultseppunct}\relax
\EndOfBibitem
\bibitem[Wolfowicz \latin{et~al.}(2013)Wolfowicz, Tyryshkin, George, Riemann, Abrosimov, Becker, Pohl, Thewalt, Lyon, and Morton]{wolfowiczAtomicClockTransitions2013}
Wolfowicz,~G.; Tyryshkin,~A.~M.; George,~R.~E.; Riemann,~H.; Abrosimov,~N.~V.; Becker,~P.; Pohl,~H.-J.; Thewalt,~M. L.~W.; Lyon,~S.~A.; Morton,~J. J.~L. Atomic Clock Transitions in Silicon-Based Spin Qubits. \emph{Nat. Nanotechnol.} \textbf{2013}, \emph{8}, 561--564\relax
\mciteBstWouldAddEndPuncttrue
\mciteSetBstMidEndSepPunct{\mcitedefaultmidpunct}
{\mcitedefaultendpunct}{\mcitedefaultseppunct}\relax
\EndOfBibitem
\bibitem[Zadrozny \latin{et~al.}(2015)Zadrozny, Niklas, Poluektov, and Freedman]{zadroznyMillisecondCoherenceTime2015}
Zadrozny,~J.~M.; Niklas,~J.; Poluektov,~O.~G.; Freedman,~D.~E. Millisecond {{Coherence Time}} in a {{Tunable Molecular Electronic Spin Qubit}}. \emph{ACS Cent. Sci.} \textbf{2015}, \emph{1}, 488--492\relax
\mciteBstWouldAddEndPuncttrue
\mciteSetBstMidEndSepPunct{\mcitedefaultmidpunct}
{\mcitedefaultendpunct}{\mcitedefaultseppunct}\relax
\EndOfBibitem
\bibitem[Shiddiq \latin{et~al.}(2016)Shiddiq, Komijani, Duan, {Gaita-Ari{\~n}o}, Coronado, and Hill]{shiddiqEnhancingCoherenceMolecular2016}
Shiddiq,~M.; Komijani,~D.; Duan,~Y.; {Gaita-Ari{\~n}o},~A.; Coronado,~E.; Hill,~S. Enhancing Coherence in Molecular Spin Qubits via Atomic Clock Transitions. \emph{Nature} \textbf{2016}, \emph{531}, 348--351\relax
\mciteBstWouldAddEndPuncttrue
\mciteSetBstMidEndSepPunct{\mcitedefaultmidpunct}
{\mcitedefaultendpunct}{\mcitedefaultseppunct}\relax
\EndOfBibitem
\bibitem[Collett \latin{et~al.}(2019)Collett, Ellers, Russo, Kittilstved, Timco, Winpenny, and Friedman]{collettClockTransitionCr7Mn2019}
Collett,~C.; Ellers,~K.-I.; Russo,~N.; Kittilstved,~K.; Timco,~G.; Winpenny,~R.; Friedman,~J. A {{Clock Transition}} in the {{Cr7Mn Molecular Nanomagnet}}. \emph{Magnetochemistry} \textbf{2019}, \emph{5}, 4\relax
\mciteBstWouldAddEndPuncttrue
\mciteSetBstMidEndSepPunct{\mcitedefaultmidpunct}
{\mcitedefaultendpunct}{\mcitedefaultseppunct}\relax
\EndOfBibitem
\bibitem[Kundu \latin{et~al.}(2022)Kundu, White, Moehring, Yu, Ziller, Furche, Evans, and Hill]{kundu2GHzClockTransition2022}
Kundu,~K.; White,~J. R.~K.; Moehring,~S.~A.; Yu,~J.~M.; Ziller,~J.~W.; Furche,~F.; Evans,~W.~J.; Hill,~S. A 9.2-{{GHz}} Clock Transition in a {{Lu}}({{II}}) Molecular Spin Qubit Arising from a 3,467-{{MHz}} Hyperfine Interaction. \emph{Nat. Chem.} \textbf{2022}, \emph{14}, 392--397\relax
\mciteBstWouldAddEndPuncttrue
\mciteSetBstMidEndSepPunct{\mcitedefaultmidpunct}
{\mcitedefaultendpunct}{\mcitedefaultseppunct}\relax
\EndOfBibitem
\bibitem[Collett \latin{et~al.}(2020)Collett, Santini, Carretta, and Friedman]{collettConstructingClocktransitionbasedTwoqubit2020}
Collett,~C.~A.; Santini,~P.; Carretta,~S.; Friedman,~J.~R. Constructing Clock-Transition-Based Two-Qubit Gates from Dimers of Molecular Nanomagnets. \emph{Phys. Rev. Res.} \textbf{2020}, \emph{2}, 032037\relax
\mciteBstWouldAddEndPuncttrue
\mciteSetBstMidEndSepPunct{\mcitedefaultmidpunct}
{\mcitedefaultendpunct}{\mcitedefaultseppunct}\relax
\EndOfBibitem
\bibitem[Little \latin{et~al.}(2023)Little, Mrozek, Rogers, Liu, McInnes, Bowen, Ardavan, and Winpenny]{littleExperimentalRealisationMultiqubit2023}
Little,~E.~J.; Mrozek,~J.; Rogers,~C.~J.; Liu,~J.; McInnes,~E. J.~L.; Bowen,~A.~M.; Ardavan,~A.; Winpenny,~R. E.~P. Experimental Realisation of Multi-Qubit Gates Using Electron Paramagnetic Resonance. \emph{Nat. Commun.} \textbf{2023}, \emph{14}, 7029\relax
\mciteBstWouldAddEndPuncttrue
\mciteSetBstMidEndSepPunct{\mcitedefaultmidpunct}
{\mcitedefaultendpunct}{\mcitedefaultseppunct}\relax
\EndOfBibitem
\bibitem[Laorenza \latin{et~al.}(2021)Laorenza, Kairalapova, Bayliss, Goldzak, Greene, Weiss, Deb, Mintun, Collins, Awschalom, Berkelbach, and Freedman]{laorenzaTunableCrMolecular2021}
Laorenza,~D.~W.; Kairalapova,~A.; Bayliss,~S.~L.; Goldzak,~T.; Greene,~S.~M.; Weiss,~L.~R.; Deb,~P.; Mintun,~P.~J.; Collins,~K.~A.; Awschalom,~D.~D.; Berkelbach,~T.~C.; Freedman,~D.~E. Tunable {{Cr}} {\textsuperscript{4+}} {{Molecular Color Centers}}. \emph{J. Am. Chem. Soc.} \textbf{2021}, \emph{143}, 21350--21363\relax
\mciteBstWouldAddEndPuncttrue
\mciteSetBstMidEndSepPunct{\mcitedefaultmidpunct}
{\mcitedefaultendpunct}{\mcitedefaultseppunct}\relax
\EndOfBibitem
\bibitem[Bayliss \latin{et~al.}(2022)Bayliss, Deb, Laorenza, Onizhuk, Galli, Freedman, and Awschalom]{baylissEnhancingSpinCoherence2022}
Bayliss,~S.~L.; Deb,~P.; Laorenza,~D.~W.; Onizhuk,~M.; Galli,~G.; Freedman,~D.~E.; Awschalom,~D.~D. Enhancing {{Spin Coherence}} in {{Optically Addressable Molecular Qubits}} through {{Host-Matrix Control}}. \emph{Phys. Rev. X} \textbf{2022}, \emph{12}, 031028\relax
\mciteBstWouldAddEndPuncttrue
\mciteSetBstMidEndSepPunct{\mcitedefaultmidpunct}
{\mcitedefaultendpunct}{\mcitedefaultseppunct}\relax
\EndOfBibitem
\bibitem[Billaud \latin{et~al.}(2023)Billaud, Balembois, Le~Dantec, Ran{\v c}i{\'c}, Albertinale, Bertaina, Chaneli{\`e}re, Goldner, Est{\`e}ve, Vion, Bertet, and Flurin]{billaudMicrowaveFluorescenceDetection2023}
Billaud,~E.; Balembois,~L.; Le~Dantec,~M.; Ran{\v c}i{\'c},~M.; Albertinale,~E.; Bertaina,~S.; Chaneli{\`e}re,~T.; Goldner,~P.; Est{\`e}ve,~D.; Vion,~D.; Bertet,~P.; Flurin,~E. Microwave {{Fluorescence Detection}} of {{Spin Echoes}}. \emph{Phys. Rev. Lett.} \textbf{2023}, \emph{131}, 100804\relax
\mciteBstWouldAddEndPuncttrue
\mciteSetBstMidEndSepPunct{\mcitedefaultmidpunct}
{\mcitedefaultendpunct}{\mcitedefaultseppunct}\relax
\EndOfBibitem
\bibitem[Mullin \latin{et~al.}(2023)Mullin, Laorenza, Freedman, and Rondinelli]{mullinQuantumSensingMagnetic2023}
Mullin,~K.~R.; Laorenza,~D.~W.; Freedman,~D.~E.; Rondinelli,~J.~M. Quantum Sensing of Magnetic Fields with Molecular Color Centers. \emph{Phys. Rev. Res.} \textbf{2023}, \emph{5}, L042023\relax
\mciteBstWouldAddEndPuncttrue
\mciteSetBstMidEndSepPunct{\mcitedefaultmidpunct}
{\mcitedefaultendpunct}{\mcitedefaultseppunct}\relax
\EndOfBibitem
\bibitem[Morton and Bertet(2018)Morton, and Bertet]{mortonStoringQuantumInformation2018}
Morton,~J.~J.; Bertet,~P. Storing Quantum Information in Spins and High-Sensitivity {{ESR}}. \emph{J. Magn. Reson.} \textbf{2018}, \emph{287}, 128--139\relax
\mciteBstWouldAddEndPuncttrue
\mciteSetBstMidEndSepPunct{\mcitedefaultmidpunct}
{\mcitedefaultendpunct}{\mcitedefaultseppunct}\relax
\EndOfBibitem
\bibitem[Kubo \latin{et~al.}(2011)Kubo, Grezes, Dewes, Umeda, Isoya, Sumiya, Morishita, Abe, Onoda, Ohshima, Jacques, Dr{\'e}au, Roch, Diniz, Auffeves, Vion, Esteve, and Bertet]{kuboHybridQuantumCircuit2011}
Kubo,~Y. \latin{et~al.}  Hybrid {{Quantum Circuit}} with a {{Superconducting Qubit Coupled}} to a {{Spin Ensemble}}. \emph{Phys. Rev. Lett.} \textbf{2011}, \emph{107}, 220501\relax
\mciteBstWouldAddEndPuncttrue
\mciteSetBstMidEndSepPunct{\mcitedefaultmidpunct}
{\mcitedefaultendpunct}{\mcitedefaultseppunct}\relax
\EndOfBibitem
\bibitem[Stephens(1976)]{stephensIntrinsicLowtemperatureThermal1976}
Stephens,~R.~B. Intrinsic Low-Temperature Thermal Properties of Glasses. \emph{Phys. Rev. B} \textbf{1976}, \emph{13}, 852--865\relax
\mciteBstWouldAddEndPuncttrue
\mciteSetBstMidEndSepPunct{\mcitedefaultmidpunct}
{\mcitedefaultendpunct}{\mcitedefaultseppunct}\relax
\EndOfBibitem
\bibitem[Griscom(1991)]{griscomOpticalPropertiesStructure1991}
Griscom,~D.~L. Optical {{Properties}} and {{Structure}} of {{Defects}} in {{Silica Glass}}. \emph{J. Ceram. Soc. Japan} \textbf{1991}, \emph{99}, 923--942\relax
\mciteBstWouldAddEndPuncttrue
\mciteSetBstMidEndSepPunct{\mcitedefaultmidpunct}
{\mcitedefaultendpunct}{\mcitedefaultseppunct}\relax
\EndOfBibitem
\bibitem[Jug \latin{et~al.}(2014)Jug, Paliienko, and Bonfanti]{jugGlassyStateMagnetically2014}
Jug,~G.; Paliienko,~M.; Bonfanti,~S. The Glassy State --- {{Magnetically}} Viewed from the Frozen End. \emph{J. Non-Cryst. Solids} \textbf{2014}, \emph{401}, 66--72\relax
\mciteBstWouldAddEndPuncttrue
\mciteSetBstMidEndSepPunct{\mcitedefaultmidpunct}
{\mcitedefaultendpunct}{\mcitedefaultseppunct}\relax
\EndOfBibitem
\bibitem[Jug and Recchia(2021)Jug, and Recchia]{jugRevealingIntrinsicMagnetism2021}
Jug,~G.; Recchia,~S. Revealing the {{Intrinsic Magnetism}} of {{Non-Magnetic Glasses}}. 2021\relax
\mciteBstWouldAddEndPuncttrue
\mciteSetBstMidEndSepPunct{\mcitedefaultmidpunct}
{\mcitedefaultendpunct}{\mcitedefaultseppunct}\relax
\EndOfBibitem
\bibitem[Weeks(1994)]{weeksManyVarietiesCenters1994}
Weeks,~R.~A. The Many Varieties of {{E}}{$\prime$} Centers: A Review. \emph{Journal of Non-Crystalline Solids} \textbf{1994}, \emph{179}, 1--9\relax
\mciteBstWouldAddEndPuncttrue
\mciteSetBstMidEndSepPunct{\mcitedefaultmidpunct}
{\mcitedefaultendpunct}{\mcitedefaultseppunct}\relax
\EndOfBibitem
\bibitem[Magruder \latin{et~al.}(2008)Magruder, Stesmans, Weeks, and Weller]{magruderEffectImplantingBoron2008}
Magruder,~R.~H.; Stesmans,~A.; Weeks,~R.~A.; Weller,~R.~A. The Effect of Implanting Boron on the Optical Absorption and Electron Paramagnetic Resonance Spectra of Silica. \emph{J. Appl. Phys.} \textbf{2008}, \emph{104}, 054110\relax
\mciteBstWouldAddEndPuncttrue
\mciteSetBstMidEndSepPunct{\mcitedefaultmidpunct}
{\mcitedefaultendpunct}{\mcitedefaultseppunct}\relax
\EndOfBibitem
\bibitem[Wang \latin{et~al.}(2020)Wang, Zhang, Sun, Du, Guan, Peng, and Wang]{wangGIrradiationEffectsBorosilicate2020}
Wang,~T.; Zhang,~X.; Sun,~M.; Du,~X.; Guan,~M.; Peng,~H.; Wang,~T. {$\gamma$}-{{Irradiation}} Effects in Borosilicate Glass Studied by {{EPR}} and {{UV}}--{{Vis}} Spectroscopies. \emph{Nucl. Instrum. Meth. B} \textbf{2020}, \emph{464}, 106--110\relax
\mciteBstWouldAddEndPuncttrue
\mciteSetBstMidEndSepPunct{\mcitedefaultmidpunct}
{\mcitedefaultendpunct}{\mcitedefaultseppunct}\relax
\EndOfBibitem
\bibitem[Kimmel and Shluger(2009)Kimmel, and Shluger]{kimmelDoublyPositivelyCharged2009}
Kimmel,~A.~V.; Shluger,~A.~L. Doubly Positively Charged Oxygen Vacancies in Silica Glass. \emph{J. Non-Cryst. Solids} \textbf{2009}, \emph{355}, 1103--1106\relax
\mciteBstWouldAddEndPuncttrue
\mciteSetBstMidEndSepPunct{\mcitedefaultmidpunct}
{\mcitedefaultendpunct}{\mcitedefaultseppunct}\relax
\EndOfBibitem
\bibitem[Kimmel \latin{et~al.}(2009)Kimmel, Sushko, Shluger, and Bersuker]{kimmelPositiveNegativeOxygen2009}
Kimmel,~A.; Sushko,~P.; Shluger,~A.; Bersuker,~G. Positive and {{Negative Oxygen Vacancies}} in {{Amorphous Silica}}. \emph{ECS Trans.} \textbf{2009}, \emph{19}, 3--17\relax
\mciteBstWouldAddEndPuncttrue
\mciteSetBstMidEndSepPunct{\mcitedefaultmidpunct}
{\mcitedefaultendpunct}{\mcitedefaultseppunct}\relax
\EndOfBibitem
\bibitem[Deligiannakis \latin{et~al.}(1998)Deligiannakis, Astrakas, Kordas, and Smith]{deligiannakisElectronicStructureGlass1998}
Deligiannakis,~Y.; Astrakas,~L.; Kordas,~G.; Smith,~R.~A. Electronic Structure of {{B}} 2 {{O}} 3 Glass Studied by One- and Two-Dimensional Electron-Spin-Echo Envelope Modulation Spectroscopy. \emph{Phys. Rev. B} \textbf{1998}, \emph{58}, 11420--11434\relax
\mciteBstWouldAddEndPuncttrue
\mciteSetBstMidEndSepPunct{\mcitedefaultmidpunct}
{\mcitedefaultendpunct}{\mcitedefaultseppunct}\relax
\EndOfBibitem
\bibitem[Ludwig \latin{et~al.}(2003)Ludwig, Nagel, Hunklinger, and Enss]{ludwigMagneticFieldDependent2003}
Ludwig,~S.; Nagel,~P.; Hunklinger,~S.; Enss,~C. Magnetic {{Field Dependent Coherent Polarization Echoes}} in {{Glasses}}. \emph{J. Low Temp. Phys.} \textbf{2003}, \emph{131}, 88--111\relax
\mciteBstWouldAddEndPuncttrue
\mciteSetBstMidEndSepPunct{\mcitedefaultmidpunct}
{\mcitedefaultendpunct}{\mcitedefaultseppunct}\relax
\EndOfBibitem
\bibitem[Nagel \latin{et~al.}(2004)Nagel, Fleischmann, Hunklinger, and Enss]{nagelNovelIsotopeEffects2004}
Nagel,~P.; Fleischmann,~A.; Hunklinger,~S.; Enss,~C. Novel {{Isotope Effects Observed}} in {{Polarization Echo Experiments}} in {{Glasses}}. \emph{Phys. Rev. Lett.} \textbf{2004}, \emph{92}, 245511\relax
\mciteBstWouldAddEndPuncttrue
\mciteSetBstMidEndSepPunct{\mcitedefaultmidpunct}
{\mcitedefaultendpunct}{\mcitedefaultseppunct}\relax
\EndOfBibitem
\bibitem[Froncisz and Hyde(1982)Froncisz, and Hyde]{fronciszLoopgapResonatorNew1982}
Froncisz,~W.; Hyde,~J.~S. The Loop-Gap Resonator: A New Microwave Lumped Circuit {{ESR}} Sample Structure. \emph{J. Magn. Reson.} \textbf{1982}, \emph{47}, 515--521\relax
\mciteBstWouldAddEndPuncttrue
\mciteSetBstMidEndSepPunct{\mcitedefaultmidpunct}
{\mcitedefaultendpunct}{\mcitedefaultseppunct}\relax
\EndOfBibitem
\bibitem[Eisenach \latin{et~al.}(2018)Eisenach, Barry, Pham, Rojas, Englund, and Braje]{eisenachBroadbandLoopGap2018}
Eisenach,~E.~R.; Barry,~J.~F.; Pham,~L.~M.; Rojas,~R.~G.; Englund,~D.~R.; Braje,~D.~A. Broadband Loop Gap Resonator for Nitrogen Vacancy Centers in Diamond. \emph{Rev. Sci. Instrum.} \textbf{2018}, \emph{89}, 094705\relax
\mciteBstWouldAddEndPuncttrue
\mciteSetBstMidEndSepPunct{\mcitedefaultmidpunct}
{\mcitedefaultendpunct}{\mcitedefaultseppunct}\relax
\EndOfBibitem
\bibitem[Joshi \latin{et~al.}(2020)Joshi, Kubasek, Nikolov, Sheehan, Costa, All{\~a}o~Cassaro, and Friedman]{joshiAdjustableCouplingSitu2020}
Joshi,~G.; Kubasek,~J.; Nikolov,~I.; Sheehan,~B.; Costa,~T.~A.; All{\~a}o~Cassaro,~R.~A.; Friedman,~J.~R. Adjustable Coupling and {\emph{in Situ}} Variable Frequency Electron Paramagnetic Resonance Probe with Loop-Gap Resonators for Spectroscopy up to {{X-band}}. \emph{Rev. Sci. Instrum.} \textbf{2020}, \emph{91}, 023104\relax
\mciteBstWouldAddEndPuncttrue
\mciteSetBstMidEndSepPunct{\mcitedefaultmidpunct}
{\mcitedefaultendpunct}{\mcitedefaultseppunct}\relax
\EndOfBibitem
\bibitem[Skuja \latin{et~al.}(2005)Skuja, Hirano, Hosono, and Kajihara]{skujaDefectsOxideGlasses2005}
Skuja,~L.; Hirano,~M.; Hosono,~H.; Kajihara,~K. Defects in Oxide Glasses. \emph{Phys. Stat. Sol. C} \textbf{2005}, \emph{2}, 15--24\relax
\mciteBstWouldAddEndPuncttrue
\mciteSetBstMidEndSepPunct{\mcitedefaultmidpunct}
{\mcitedefaultendpunct}{\mcitedefaultseppunct}\relax
\EndOfBibitem
\bibitem[Muruganathan and Mizuta(2021)Muruganathan, and Mizuta]{muruganathanBoronVacancyColor2021}
Muruganathan,~M.; Mizuta,~H. Boron Vacancy Color Center in Diamond: {{Ab}} Initio Study. \emph{DRM} \textbf{2021}, \emph{114}, 108341\relax
\mciteBstWouldAddEndPuncttrue
\mciteSetBstMidEndSepPunct{\mcitedefaultmidpunct}
{\mcitedefaultendpunct}{\mcitedefaultseppunct}\relax
\EndOfBibitem
\bibitem[Umeda \latin{et~al.}(2022)Umeda, Watanabe, Hara, Sumiya, Onoda, Uedono, Chuprina, Siyushev, Jelezko, Wrachtrup, and Isoya]{umedaNegativelyChargedBoron2022}
Umeda,~T.; Watanabe,~K.; Hara,~H.; Sumiya,~H.; Onoda,~S.; Uedono,~A.; Chuprina,~I.; Siyushev,~P.; Jelezko,~F.; Wrachtrup,~J.; Isoya,~J. Negatively Charged Boron Vacancy Center in Diamond. \emph{Phys. Rev. B} \textbf{2022}, \emph{105}, 165201\relax
\mciteBstWouldAddEndPuncttrue
\mciteSetBstMidEndSepPunct{\mcitedefaultmidpunct}
{\mcitedefaultendpunct}{\mcitedefaultseppunct}\relax
\EndOfBibitem
\end{mcitethebibliography}


\providecommand{\latin}[1]{#1}
\makeatletter
\providecommand{\doi}
  {\begingroup\let\do\@makeother\dospecials
  \catcode`\{=1 \catcode`\}=2 \doi@aux}
\providecommand{\doi@aux}[1]{\endgroup\texttt{#1}}
\makeatother
\providecommand*\mcitethebibliography{\thebibliography}
\csname @ifundefined\endcsname{endmcitethebibliography}  {\let\endmcitethebibliography\endthebibliography}{}
\begin{mcitethebibliography}{2}
\providecommand*\natexlab[1]{#1}
\providecommand*\mciteSetBstSublistMode[1]{}
\providecommand*\mciteSetBstMaxWidthForm[2]{}
\providecommand*\mciteBstWouldAddEndPuncttrue
  {\def\EndOfBibitem{\unskip.}}
\providecommand*\mciteBstWouldAddEndPunctfalse
  {\let\EndOfBibitem\relax}
\providecommand*\mciteSetBstMidEndSepPunct[3]{}
\providecommand*\mciteSetBstSublistLabelBeginEnd[3]{}
\providecommand*\EndOfBibitem{}
\mciteSetBstSublistMode{f}
\mciteSetBstMaxWidthForm{subitem}{(\alph{mcitesubitemcount})}
\mciteSetBstSublistLabelBeginEnd
  {\mcitemaxwidthsubitemform\space}
  {\relax}
  {\relax}

\bibitem[Joshi \latin{et~al.}(2020)Joshi, Kubasek, Nikolov, Sheehan, Costa, All{\~a}o~Cassaro, and Friedman]{joshiAdjustableCouplingSitu2020}
Joshi,~G.; Kubasek,~J.; Nikolov,~I.; Sheehan,~B.; Costa,~T.~A.; All{\~a}o~Cassaro,~R.~A.; Friedman,~J.~R. Adjustable Coupling and {\emph{in Situ}} Variable Frequency Electron Paramagnetic Resonance Probe with Loop-Gap Resonators for Spectroscopy up to {{X-band}}. \emph{Rev. Sci. Instrum.} \textbf{2020}, \emph{91}, 023104\relax
\mciteBstWouldAddEndPuncttrue
\mciteSetBstMidEndSepPunct{\mcitedefaultmidpunct}
{\mcitedefaultendpunct}{\mcitedefaultseppunct}\relax
\EndOfBibitem
\end{mcitethebibliography}

\end{document}


\section{Additional Data}

\begin{figure}[ht!]
 \centering
 \includegraphics[width=1\linewidth]{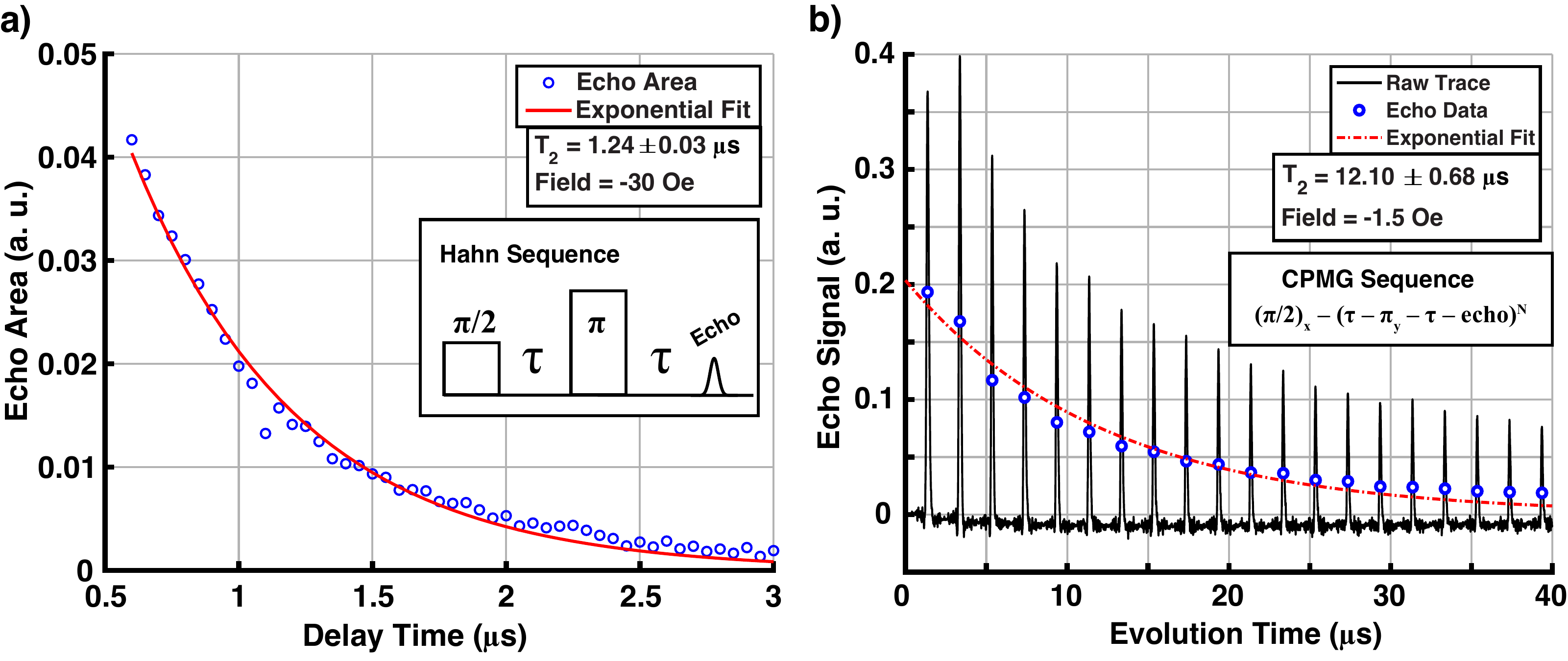}
 \caption{\textbf{Coherence pulse sequence examples.} (a) Borosilicate glass coherence time is measured via the Hahn sequence at $-30$~Oe, 5117~MHz, and 1.8~K. Blue shows integrated echo area, and red shows a best-fit exponential decay. The inset illustrates the Hahn pulse sequence -- to measure coherence, the delay time $\uptau$ between pulses is varied. (b) Borosilicate glass coherence measured via the CPMG pulse sequence at $-1.5$~Oe, 5117~MHz, and 1.8~K. Black data is a raw CPMG trace; echo peaks are spaced by $2\uptau$. Blue shows the integrated echo area, and red shows a best-fit exponential decay. The inset shows a CPMG pulse sequence as used in this study.}
 \label{fig:coherence}
\end{figure}

We examined each spin system presented in the main manuscript using ESR techniques to characterize the clock-transition behavior. Fig.~\ref{fig:coherence}(a) shows an example  Hahn echo decay -- the delay time $\uptau$ is varied between pulses, allowing for echoes with differing decoherence to be measured. Fig.~\ref{fig:coherence}(b) shows a similar experiment, except with the CPMG pulse sequence. By repeatedly applying refocusing pulses as prescribed by the sequence $(\uppi/2)_x - (\tau - \uppi_y - \tau - \text{echo})^N$, coherence can be enhanced.

\begin{figure}[ht!]
 \centering
 \includegraphics[width=1\linewidth]{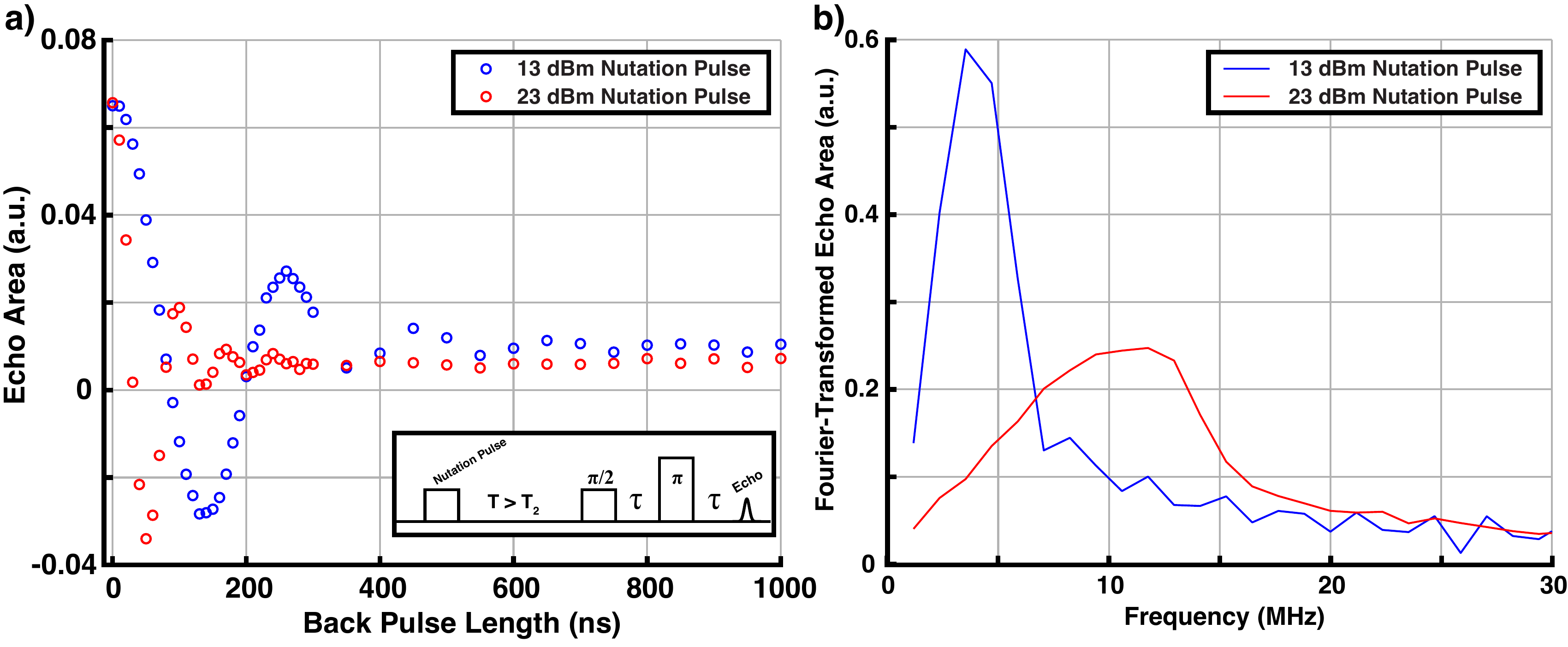}
 \caption{\textbf{Rabi oscillations in borosilicate glass at the clock transition.} (a) Two examples of driven Rabi oscillations; in blue, the nutation pulse is amplitude-matched to the $\uppi/2$ pulse (13~dBm) and the length of time of the nutation pulse is varied; in red, the nutation pulse is set to have maximum amplitude (23~dBm) so as to maximize the Rabi frequency. The pulse sequence is shown in the inset of (a). (b) Fourier-transformed Rabi oscillations of the data shown in (a); Rabi frequency peaks can be seen at $\sim3$~MHz (blue) and $\sim10$~MHz (red), though both peaks have substantial breadth and the high-power peak is much broader. The DC point in the Fourier-transformed data has been excluded for clarity. Experiments were performed at 0~Oe, 1.8~K, and 5116~MHz; in both curves the time $T$ between the nutation pulse and the $\uppi/2$ pulse was 20~$\upmu$s and $\uptau = 1000$~ns.}
 \label{fig:Rabi_boro}
\end{figure}

To demonstrate coherent control of the two-level system, we drove Rabi oscillations at the CT in each system we studied. Fig.~\ref{fig:Rabi_boro}(a) shows two examples of Rabi oscillations in borosilicate glass at 0~Oe, 1.8~K, and 5116~MHz. The pulse sequence for this experiment is shown in the inset on Fig.~\ref{fig:Rabi_boro}(a). The first pulse in this sequence is a nutation pulse that we vary in duration to characterize the Rabi dynamics. The data in blue shows the results of a Rabi experiment wherein the nutation pulse was tuned such that its amplitude matched the amplitude of the $\uppi/2$ pulse (13~dBm). Increasing the power of the nutation pulse to our experimental maximum (23~dBm, red), produces faster oscillations.   Fig.~\ref{fig:Rabi_boro}(b) shows the Fourier Transform of the data in (a) with each data set having a well-defined peak in the frequency domain. Since the Rabi frequency should scale as $\sqrt{P}$, where $P$ is the radiation power, we expect our two Rabi frequencies to have a ratio of $\sqrt{10}$. The data in Fig.~\ref{fig:Rabi_boro}(b) is consistent with this prediction. There is some breadth to each peak in (b) related to the decay in the oscillations. We observe that a higher-power nutation pulse drives the spins faster, but generates a faster decay and a correspondingly broader distribution of frequencies.

\begin{figure}[ht!]
 \centering
 \includegraphics[width=0.65\linewidth]{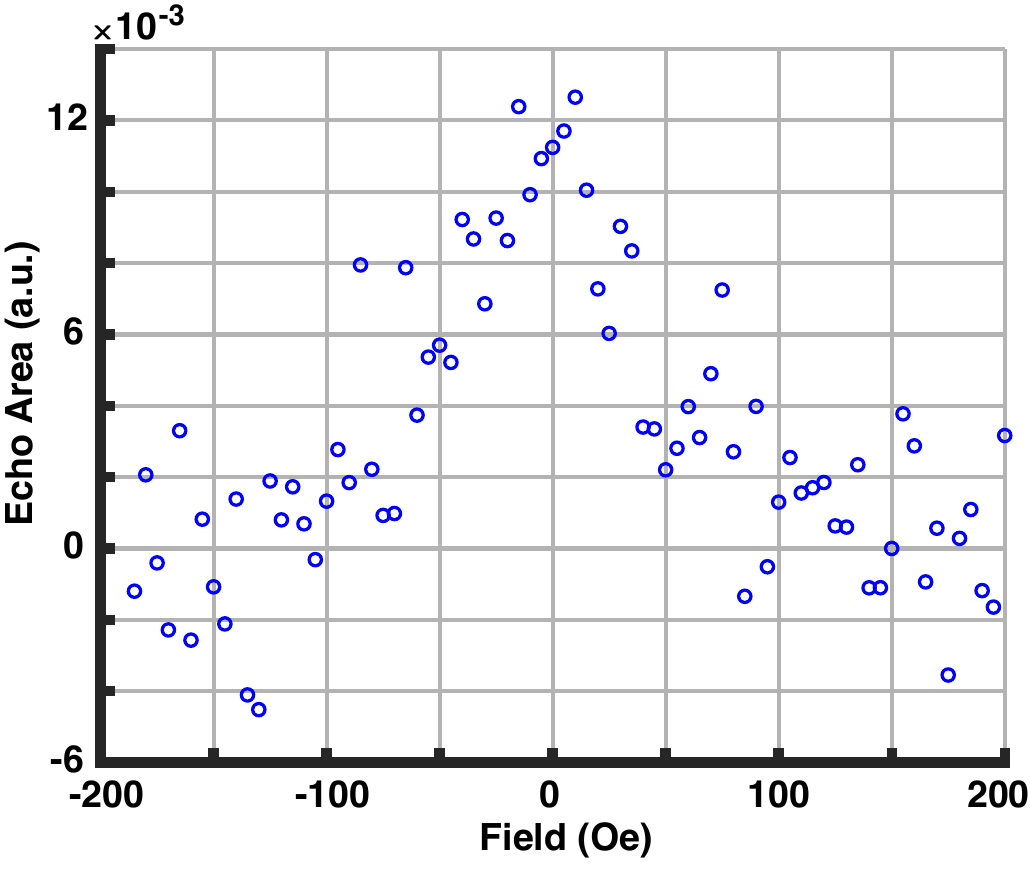}
 \caption{\textbf{EDFS of Vycor 7913 at 3757~MHz and 1.8~K.} Vycor  shows an increase in echo area at zero field, when probed with a Hahn echo sequence with the delay time $\uptau = 600$~ns.}
 \label{fig:vycorCT}
\end{figure}

\begin{figure}[ht!]
 \centering
 \includegraphics[width=1\linewidth]{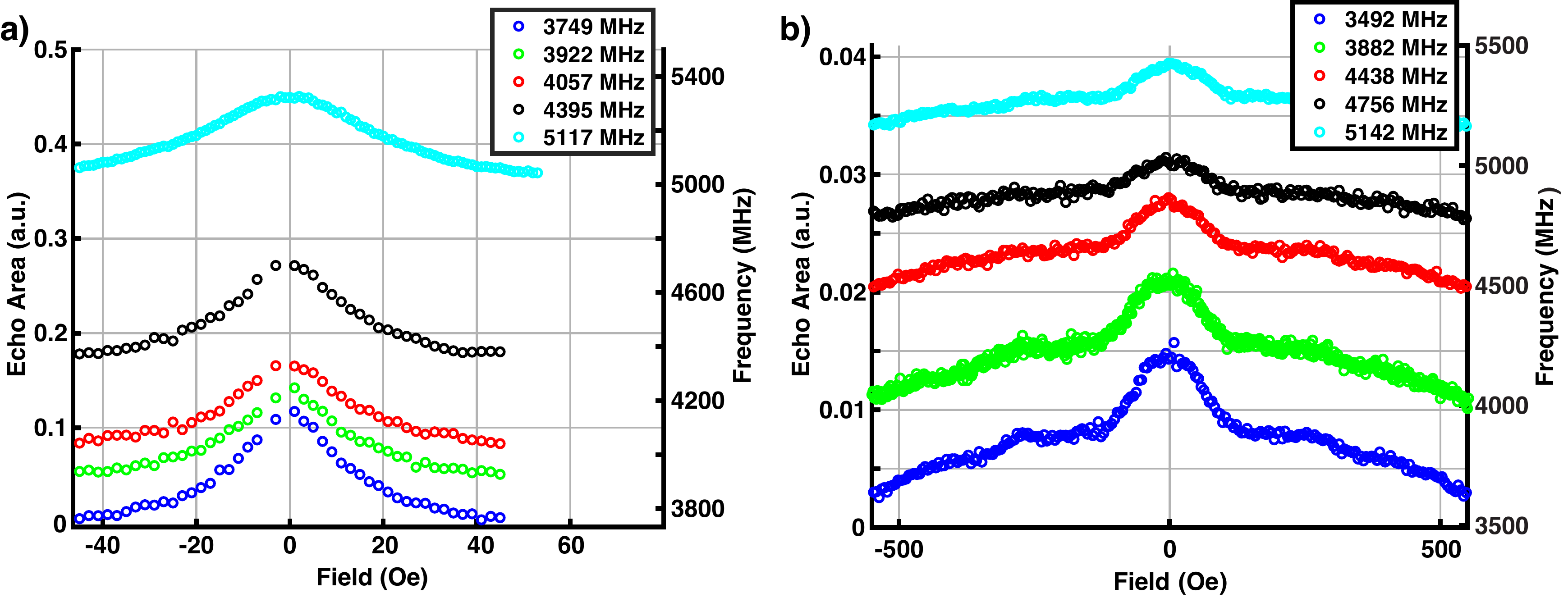}
 \caption{\textbf{Waterfall plots of EDFS of borosilicate and AGC40.} (a) A waterfall plot of the zero-field increase in echo signal for borosilicate glass, shown across five frequencies. Within the range 3.5 -- 5.5~GHz the echo signal is strong. Below this frequency range, echo signal falls off rapidly and becomes undetectable. (b) A waterfall plot of the zero-field increase in echo signal for AGC40 glass, shown across five frequencies. Similarly, in the 3.5 -- 5.5~GHz range, the echo signal is strong and below this range, echo signal falls off rapidly and becomes undetectable. In each panel, each EDFS spectrum is shifted vertically such that he rightmost data point corresponds to the ESR frequency at which the data was taken, indicated in the right-hand y-axis. All experiments were performed at 1.8~K.}
 \label{fig:EDFSwaterfall}
\end{figure}

Fig.~\ref{fig:vycorCT} shows an echo-detected field stepped (EDFS) trace for Corning Vycor 7913, a porous borosilicate glass, which, like the other glasses in Table~1 in the main manuscript, shows an increase in echo signal at zero field when probed with a Hahn echo sequence. Unlike other samples we investigated, the echo signal at 1.8~K is extremely small, making characterization of coherence in Vycor 7913 difficult. After annealing this sample, we were no longer able to detect an echo signal.

Fig.~1(b) of the main manuscript shows an inhomogeneously broadened avoided level crossing diagram to account for the range of frequencies (3.5 -- 5.5~GHz) over which the CT in each material is observed. So long as one is working with an ESR frequency in that range, there is no significant frequency dependence in the signal amplitude or spin dynamics for any sample. Below 3.5~GHz, the magnitude of the echo signal rapidly falls off; we could not explore frequencies above 5.5~GHz with the apparatus used. Fig.~\ref{fig:EDFSwaterfall} provides two examples of sample frequency dependence, showing similar behavior in EDFS spectra across a broad range of frequencies. By taking advantage of the \textit{in-situ} frequency and coupling adjustment of our ESR probe, each sample could be examined over a wide range of frequencies without changing microwave resonator~\cite{joshiAdjustableCouplingSitu2020}. Fig.~\ref{fig:EDFSwaterfall}(a) shows a waterfall plot of EDFS spectra for borosilicate glass at 1.8~K. For each of the five frequencies shown, a substantial increase in echo signal occurs at zero field. Each EDFS spectrum was acquired using a standard Hahn echo sequence. With the exception of the 5117~MHz spectrum, which used a delay time of $\uptau = 1000$~ns, all spectra were acquired with $\uptau = 1200$~ns. The spectrum recorded at 5117 MHz shows a broader CT than the lower-frequency spectra ( $\sim50\%$ broader, as measured by the FWHM of the CT). Each EDFS spectrum is shifted vertically such that the rightmost data point corresponds to the ESR frequency at which the data was taken, indicated in the right-hand y-axis. Fig.~\ref{fig:EDFSwaterfall}(b) shows a similar waterfall plot of EDFS spectra for AGC40 glass at 1.8~K. All spectra in Fig.~\ref{fig:EDFSwaterfall}(b) were measured using $\uptau = 1000$~ns. In both cases, the magnitude of the CT peak is similar across a broad range of frequencies.

\begin{table*}
\begin{tabular}{c||c|c|c|c|c}
 ~ & Quartz & Fused Silica & Alumina & B$_2$O$_3$ & Glass 0100 (Soda Lime) \\
 \hline
 SiO$_2$ (\%) & 99.99 & 99.99 & 0 & 0 & 69 \\
 B$_2$O$_3$ (\%) & 0 & 0 & 0 & 99.9 & 1 \\
 Al$_2$O$_3$ (\%) & 0 & 0 & 99.5 & 0 & 4 \\
 Na$_2$O (\%) & 0 & 0 & 0 & 0 & 13 \\
 K$_2$O (\%) & 0 & 0 & 0 & 0 & 3 \\
 CaO (\%) & 0 & 0 & 0 & 0 & 5 \\
 BaO (\%) & 0 & 0 & 0 & 0 & 2 \\
 MgO (\%) & 0 & 0 & 0 & 0 & 3 \\
 \hline
\end{tabular}
\caption{\textbf{Mass percentage composition of each sample studied in which no zero-field Hahn echo was detected.} All data was sourced from commercially available descriptions of the material from the manufacturer.}
\label{tab:noCTsamples}
\end{table*}

We compared glass samples with clock transitions to stoichiometrically similar samples in which we found no evidence of a zero-field Hahn echo. Specific elemental data by mass is shown in Table~\ref{tab:noCTsamples}. Neither quartz (crystalline SiO$_2$) nor fused silica showed any detectable echo near zero field. Similarly, boron trioxide (B$_2$O$_3$) and alumina 
(Al$_2$O$_3$) showed no zero-field signal. Soda lime glass, which contains numerous interstitial impurities (in particular, Na and Ca) but few substitutional defects, also showed no zero-field signal. The sourcing of materials in Table~\ref{tab:noCTsamples} is as follows: quartz and fused silica were sourced as wafers from MSE Supplies; alumina (PubChem Substance ID:
329752862) and boron trioxide (PubChem Substance ID: 24866686) were sourced as powders from Sigma Aldrich; and the soda lime glass was sourced from Globe Scientific.

\begin{table*}
\begin{tabular}{c||c|c|c|c|c}
 ~ & Borosilicate & Duran & Aluminosilicate 0812 & Vycor 7913 & AGC40 \\
 \hline
 Temperature ($^\text{o}$C) & 565 & 565 & 565 & 1040 & 1040 \\
 Time (minutes) & 30 & 30 & 30 & 30 & 30 \\
 Atmosphere & Air & Air & Air & Air & Air \\
 CT Visible? & Yes & Yes & No & No & No \\
 \hline
\end{tabular}
\caption{\textbf{Annealing process for each sample.} Each sample that showed a CT was annealed according to published annealing temperatures in manufacturer-provided datasheets. Annealing was performed in air at atmospheric pressure for 30 minutes for all samples, at the University of Massachusetts Amherst Scientific Glassblowing Laboratory. The final row in the table indicates whether the sample exhibited a zero-field Hahn echo signal after annealing.}
\label{tab:annealing}
\end{table*}

Each sample that demonstrated a CT was subsequently annealed to examine whether the CT signal is consistent with the presence of vacancies. Table~\ref{tab:annealing} indicates the annealing conditions for each sample and whether it demonstrated echo after annealing. All samples were annealed at temperatures according to the specified annealing temperature provided by the material's manufacturer. The borosilicate sample demonstrated significantly reduced echo signal and increased coherence at the CT after annealing, as discussed in Fig.~5 of the main manuscript. In contrast, Duran showed no significant change in spin dynamics after annealing. Aluminosilicate 0812 glass, Vycor 7913, and AGC40 produced no zero-field signal after annealing. With the notable exception of the Duran sample, these observations are consistent with the hypothesis that the CT in each sample is generated, at least in part, by the presence of vacancies in the material; if annealing substantially reduces the number of vacancies in the sample, one would expect a reduction in the signal from spin centers containing vacancies. We note that annealing may cause other structural changes in the material and there may be other hypotheses that are consistent with our observations. 

\bibliography{Silica_Defects_and_Clock_Transitionssup}